\def\year{2020}\relax
\documentclass[letterpaper]{article} 
\usepackage{aaai20}  
\usepackage{times}  
\usepackage{helvet} 
\usepackage{courier}  
\usepackage[hyphens]{url}  
\usepackage{graphicx} 
\usepackage{graphicx}  
\usepackage{booktabs}

\title{Top Comment or Flop Comment?\\ Predicting and Explaining User Engagement in Online News Discussions}

\author{Julian Risch, Ralf Krestel\\
Hasso Plattner Institute, University of Potsdam, Germany\\
firstname.lastname@hpi.de
}
\begin{document}

\maketitle

\begin{abstract}
Comment sections below online news articles enjoy growing popularity among readers.
However, the overwhelming number of comments makes it infeasible for the average news consumer to read all of them and hinders engaging discussions.
Most platforms display comments in chronological order, which neglects that some of them are more relevant to users and are better conversation starters.

In this paper, we systematically analyze user engagement in the form of the upvotes and replies that a comment receives.
Based on comment texts, we train a model to distinguish comments that have either a high or low chance of receiving many upvotes and replies.
Our evaluation on user comments from \emph{TheGuardian.com} compares recurrent and convolutional neural network models, and a traditional feature-based classifier. 
Further, we investigate what makes some comments more engaging than others. 
To this end, we identify engagement triggers and arrange them in a taxonomy.
Explanation methods for neural networks reveal which input words have the strongest influence on our model's predictions.
In addition, we evaluate on a dataset of product reviews, which exhibit similar properties as user comments, such as featuring upvotes for helpfulness.
\end{abstract}

\section{User Comments in Online News Discussions}
Thirty years ago, newspapers received hand-written letters to the editor and selected maybe a handful for publication.
This was called reader engagement and was the only way for readers to interact with other readers and/or the newspaper via public discussion.
With the rise of the World Wide Web, the establishment of online news platforms, and the appearance of online discussion sections, the situation has changed drastically.
Nowadays, irrespective of who the readers are and what they think, they can exercise their right to freedom of speech and freely share their opinion.

On the flip side, the ever-increasing number of comments not only distracts readers, but also hinders engagement. 
No news consumer is able to read through all the comments.
Overwhelmed by hundreds to thousands of comments, new users give up on joining the discussion.
A current approach for coping with this information overload is to highlight comments that are especially interesting in the eyes of the editors.
This manual effort is costly and comes on top of the task of moderating hate speech and other banned content.

Major news platforms allow users to upvote comments, but for several reasons these platforms do not use votes as a ranking criterion for comments.
First, there is the cold start problem: Whenever a new comment is posted, it has not yet received any upvotes. 
An accordingly low rank affects the comment's exposure to users and reduces its chance of ever receiving any upvotes.
Moreover, such a ranking algorithm can easily be gamed.
Malicious users can register multiple accounts or collaborate to break the ranking system and upvote comments of their favored opinion.

Today's platforms refrain from using an upvote-based ranking algorithm and simply sort comments chronologically.
They give no incentive for the described manipulations.
Thus, casting an upvote conveys a sense of relevance to the respective user --- but nothing more.
A comment receives many upvotes if it motivates many users to engage by voting for it.
We make use of this information to build a dataset of comments that are either most or least engaging.
Note that some platforms also allow users to downvote comments, which we do not analyze in our work.

Voting on a comment is a rather basic way to interact.
In contrast, replying to another user's comment actually starts a conversation.
Users reply to comments for different reasons.
For example, they want to correct another user's error, give their personal view, or express consent or dissent.
While the number of upvotes reveals a comment's popularity, the same does not hold for the number of replies.
Besides upvotes, we also consider replies as a form of user engagement and give insights into their interplay.
\begin{figure}
    \centering
    \includegraphics[width=1\linewidth]{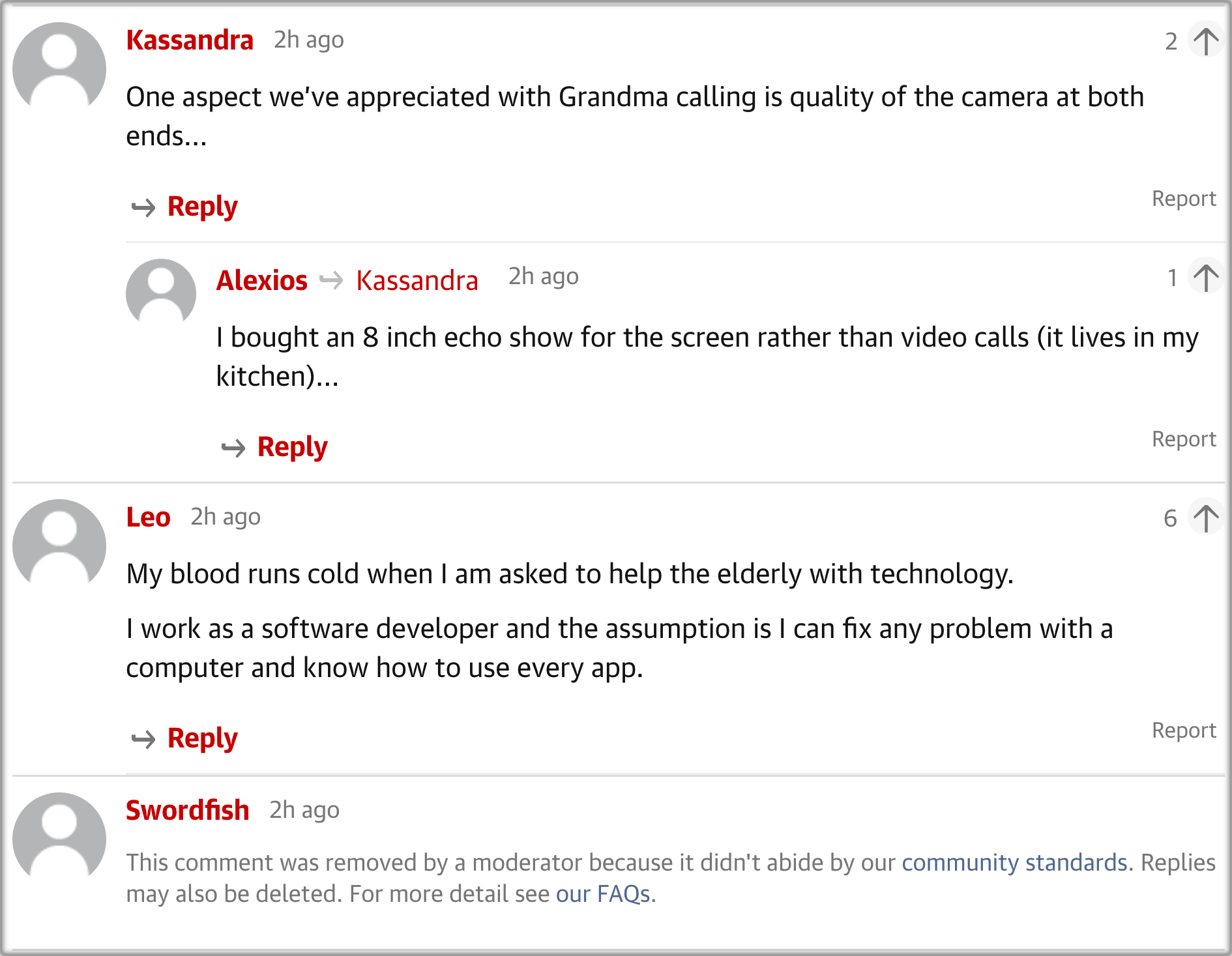}
    \caption{On \emph{TheGuardian.com}, readers can post a comment, cast an upvote, and reply to another user's comment.}
    \label{fig:screenshot}
\end{figure}

In this paper, we classify engaging comments without costly manual annotation effort by editors.
Therefore, we leverage user reactions that are inherent to online discussions: comment upvotes and replies.
The number of these reactions distinguishes the most and the least engaging comments, which we also refer to as top comments and flop comments.
In our experiments we analyze a real-world dataset of user comments from the British online news platform \emph{TheGuardian.com.}
Figure~\ref{fig:screenshot} is a screenshot of the platform's comment section.
For illustration purposes, we list two comments that generated a large amount of engagement in the form of many upvotes or replies:
\begin{enumerate}
\item ``The brexiters are achieving their wish: they're turning the UK into the kind of second rate country they can feel at home in.''
2,615 upvotes, 3 replies
\item ``Can somebody please explain to me why some people are so rabidly anti-gay marriage?''
82 upvotes, 20 replies
\end{enumerate}

The first comment refers to an anticipated loss of 1,000 jobs in EU authorities located in the UK as a consequence of Brexit.
The number of received upvotes is extraordinarily high, presumably because many anti-brexiters identify with the expressed opinion.
The second comment was posted half a year before the UK parliament legalized same-sex marriage.
It was a topic of controversial discussions at that time with replies containing different opinions and address the user's request for explanation.

Being able to automatically detect such engaging comments is important for many applications.
The most obvious one would be a ranking criterion to display user comments from most engaging to least engaging.
Another application field would be in the context of recommending comments to readers, either to reply or to simple read them.
Currently this recommendation is done manually through editor picks.
Finally, users writing many engaging comments could be rewarded by the news platform.
This would further incentivize users to write high quality comments.

\paragraph{Contributions} 
In summary, the contributions of this paper are: 
(1) defining user engagement in online discussions based on adjusted number of upvotes and replies;
(2) designing a taxonomy that explains engagement triggers;
(3) proposing a neural network model to distinguish most and least engaging comments based on their text;
(4) evaluating classification accuracy of the proposed model on two datasets: user comments and product reviews.
Our implementation, the evaluation datasets, and two models of domain-specific word embeddings are published online.\footnote{\url{http://hpi.de/naumann/projects/repeatability/text-mining.html}}

\section{Related Work}
\label{sec:relatedwork}
A growing body of research aims to foster respectful and fruitful discussions on the Web.
Applications of this research manifest in real-world system implementations that support moderators and community managers, for example, by predicting how many comments a news article will receive~\cite{krestel-naacl18}, identifying comments that require moderation~\cite{Schabus2018,risch2018delete} or highlighting comments that are worth reading~\cite{park2016supporting}.
To this end, there are two primary directions of related work on comment classification: identifying either toxic or high-quality comments.

The term \emph{toxic comments} comprises hate speech, insults, threats, profanity, and content that otherwise makes users leave a discussion.
Platforms enforce a ban on toxic comments through manual moderation, but the ever-increasing number of comments renders this effort infeasible.
Several studies work towards automation of this step and train deep neural networks on large datasets of annotated comments~\cite{nobata2016abusive,wulczyn2017ex,badjatiya2017deep}.
The preparation and analysis of such datasets is a complex research task on its own~\cite{Schabus2017,chen2017presenting}.
A significant challenge is the inherent class imbalance of the data: typically, less than five percent of the comments are toxic~\cite{xu2012learning,risch2018delete}. 
Recently, it has been proposed to not only classify single comments but also predict whether the tone of a sequence of comments is getting out of hand~\cite{zhang2018}.

Highlighting high-quality comments is the complementary task to deleting toxic comments.
Related work defines the notion of quality in different ways, varying from \emph{engaging, respectful, and informative}~\cite{Napoles2017AutomaticallyIG} to \emph{interesting or thoughtful}~\cite{diakopoulos2015picking} and \emph{constructive}~\cite{kolhatkar2017using}.
Similar to the task of toxic comment classification, state-of-the-art approaches use supervised machine-learning and require large amounts of training data, e.g., 2.3k annotated conversations~\cite{napoles2017finding} or 30k annotated comments~\cite{kolhatkar2017using}.
We refrain from costly annotation efforts in our work and instead draw on information inherent to the data: upvotes and replies by users.

\citeauthor{kolhatkar2017using}~\shortcite{kolhatkar2017using} use editor picks from the \emph{New York Times} as positive training samples to learn to identify constructive comments.
These picks are a selection of comments judged as interesting or thoughtful by news editors.
Negative training samples are taken from \emph{Yahoo News} comments that were annotated as non-constructive in previous work~\cite{napoles2017finding}.
\citeauthor{lampe2004slash}~\shortcite{lampe2004slash} find that users generally agree on what comments are of high or low quality.
However, users pay more attention to the earliest comments and top-level comments in a conversation than to responses.
This finding motivates us to identify and remove this position bias in our dataset. 
Consequently, the number of upvotes and replies does not depend on the comment's position in the chronological ranking anymore.
We refer to this position as the comment's rank in the following.

Online discussions are also mined to predict popularity of news stories~\cite{rizos2016predicting}, measure how controversial a comment is~\cite{gomez2008statistical}, or rank comments by persuasiveness~\cite{wei2016post}.
\citeauthor{hsu2009ranking}~\shortcite{hsu2009ranking} make use of upvotes to rank comments, which is similar to parts of our approach.
They measure a comment's visibility (exposure to users) by considering the popularity of the corresponding news article and the time between the publication of the article and the comment.
Inspired by this idea, we use a comment's position in the chronological ranking to account for its visibility.

Related to our work, there is research on conversation modeling~\cite{kumar2010dynamics,wang2012user,gomez2013likelihood,backstrom2013characterizing,aragon2017generative,medvedev2019modelling} and on the dynamics of re-tweets~\cite{zhang2016retweet,kobayashi2016tideh}.
However, the motivation behind re-tweeting is to spread information in a social network, and in this regard it differs from replies in news discussions.
The reasons users post comments on news articles are manifold.
They range from expressing an opinion, asking questions, and correcting factual errors, to giving misinformation with the intent of seeing the community's reaction~\cite{diakopoulos2011towards}.
We propose a taxonomy to characterize the different kinds of comments that trigger engagement by other users.

\citeauthor{Berry2017}~\shortcite{Berry2017} study the ranking of posts on public Facebook pages.
They compare chronological ranking to ranking via social feedback and find that the latter has a positive effect on response quality.
This insight motivates further research on ranking criteria for online comments aside from chronological ranking.

Most related work refrains from using upvotes as a feature, because of the many different factors that influence the number of upvotes a comment receives, such as its rank.
At least, the interplay of a post's title, text, and publication time to predict user votes on Reddit and YouTube have been subject to research~\cite{lakkaraju2013s,siersdorfer2010how}.
Chronological ranking in discussion threads is an essential difference in news comments compared to, for example, posts on Twitter or Facebook that can stand alone without a conversational context.
In contrast to related work that predicts the popularity of a news article and the number of received user comments~\cite{krestel-naacl18}, we predict the users' interactions with a comment.
To this end, we neglect the news article text and focus on the comment text, upvotes, and replies.

\section{Characterizing Users and Comments}
\label{sec:datasets}
In this section, we introduce and analyze a dataset of user comments from \emph{TheGuardian.com}.
There is an inherent position bias in the number of upvotes and replies and, after removing this bias, we find that top and flop comments differ in their average length and sentiment.
Further, we visualize words that occur more often in either top or flop comments.
Last but not least, we introduce a taxonomy to systematically categorize different types of engaging comments.

\subsection{User Comment Dataset}
The dataset comprises 61 million comments posted between 2006 and 2018.
1.2 million users contributed them in discussions of 600k news articles.
For each comment, there is the comment text, user name, publication timestamp, corresponding news article, upvotes, and parent comment (if applicable).
Half of the comments (53 percent) are replies to another user's comment and thus have a reference to this parent.
Before November 2011, there was no option to post a reply in reference to another user’s comment on \emph{TheGuardian.com}~\cite{belam2011adding}.
Therefore, we limit the dataset to the time after 2011 whenever we study the replies. 
We neglect that there was another change on the platform in 2012, when a single-level threaded design was introduced for the comment section~\cite{hanman2012threading}.
\citeauthor{budak2017threading}~\shortcite{budak2017threading} analyze the impact of this change on users and their discussions.

\begin{figure}
    \centering
    \includegraphics[width=1\linewidth]{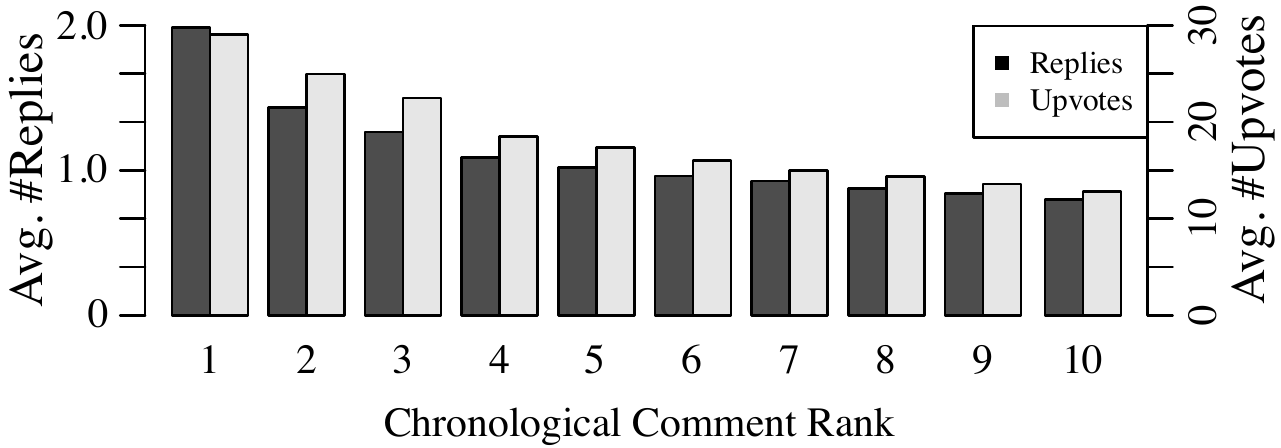}
    \caption{A comment's average number of received upvotes and replies correlates with its chronological rank in the discussion thread. This correlation is called position bias.}
    \label{fig:upvotes_and_replies_by_rank}
\end{figure}

Upvotes cover the full timespan from 2006 to 2018 so that there is no need to limit the dataset when we study them.
There are 260 million upvotes in total.
While we have no knowledge of when, why and from whom a particular comment received upvotes, we know its final number of upvotes.
We identify a position bias in the upvotes: the number of upvotes and replies that a comment receives depends on its rank.
Figure~\ref{fig:upvotes_and_replies_by_rank} visualizes this dependency and reveals that earlier comments receive more upvotes and replies on average.
In line with related work~\cite{hsu2009ranking}, we attribute the advantage of earlier comments to their greater exposure to more readers.
For this reason, the raw upvote and reply count is not enough to judge a comment's relevance to users in comparison with other comments.
That is why we propose an approach to normalize the counts and thereby prevent the position bias from distorting the results.

In short, this approach transforms the absolute counts to relative numbers and afterwards groups all comments by their rank.
For example, we compare a comment at rank 3 to all comments that appeared in other discussions at the same rank 3.
Let us assume that the comment received 20 percent of all upvotes on the ten first comments in its corresponding article discussion. 
If comments at rank 3 receive on average less than 20 percent of the upvotes, we have identified a top comment, otherwise a flop comment.
We describe the approach in more detail in the section \emph{Distinguishing Top and Flop Comments}.

Based on the approach, we distinguish between two sets of the most and least engaging (top and flop) comments and analyze their differences.
As user engagement varies by news topic~\cite{aldous2019view}, we reduce the topical variety by limiting our analysis to comments on articles in the politics section.
It is the section with the largest number of comments received.
Table~\ref{tab:analytics} compares the most and least engaging comments with regard to their average length, readability, and sentiment.
Comments that generate less engagement are on average shorter and more often have a neutral sentiment.
However, there is no difference in readability or the use of function words and personal pronouns.
We use the automated readability index (ARI) to evaluate the readability.
It is a standard metric that takes into account a text's characters per word and words per sentence.

\begin{table}
  \caption{The most and the least engaging comments differ in length and amount of neutral sentiment.}
  \label{tab:analytics}
  \centering
  \fontsize{9.0pt}{10.0pt} \selectfont 
  \begin{tabular}{lrrrr}
    \toprule
          & \multicolumn{2}{c}{Upvotes} & \multicolumn{2}{c}{Replies} \\
    \midrule
     Average per Comment       & Most & Least & Most & Least \\
    \midrule
Number of Words & 75.54 & 43.68 & 76.82 & 38.52 \\
Rate of Function Words & 0.43 & 0.43 & 0.44 & 0.43 \\
Rate of Personal Pronouns & 0.13 & 0.12 & 0.12 & 0.13 \\
Readability Index & 9.82 & 9.08 & 9.50 & 9.14 \\
Positive Sentiment & 0.47 & 0.48 & 0.49 & 0.45 \\
Neutral Sentiment & 0.07 & 0.23 & 0.09 & 0.20 \\
Negative Sentiment & 0.46 & 0.30 & 0.42 & 0.34 \\
  \bottomrule
\end{tabular}
\end{table}

\begin{figure}
    \centering
    \includegraphics[width=\columnwidth]{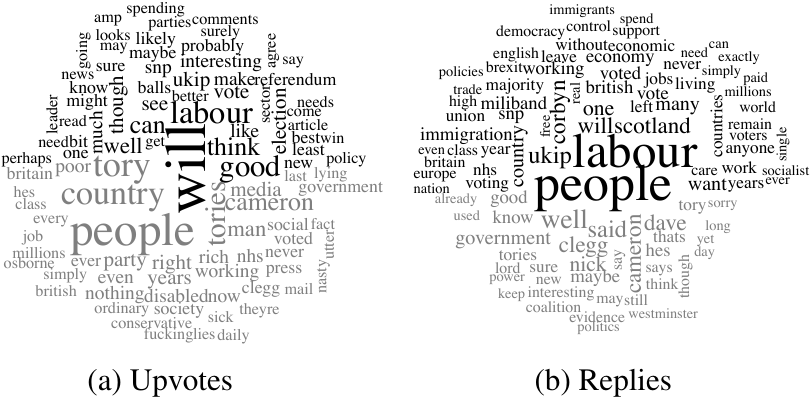}
    \caption{Comparison word clouds show indicative words for classes of the most (black) and least (grey) engaging comments. For example, comments that mention \emph{people} receive few upvotes but many replies.}
    \label{fig:comparison-wordcloud}
\end{figure}

Figure~\ref{fig:comparison-wordcloud} compares the usage of the 100 most frequent words in comments that received the most or the least upvotes or replies.
The word clouds display a word in the top half (black font) if it occurs more often in the most engaging comments and in the bottom half (gray font) if it occurs more often in the least engaging comments.
The font size corresponds to the difference in the word's relative frequencies in both classes.
For example, the relative frequency of the word \emph{Labour} is 0.39 percent in comments that receive the most replies and 0.27 percent in comments that receive the least replies.
The comparably large difference between these frequencies is illustrated by the word's large font size.

The most engaging comments mention the word \emph{Labour} more often and the word \emph{Tory} less often.
The same relation holds for politicians of the respective parties, e.g., for Jeremy Corbyn (Labour) and David Cameron (Tory).
A reason for this might be the political orientation of \emph{TheGuardian.com} readers: according to a post-election survey, 73 percent voted for the Labour party and 8 percent for the Tory party in the 2017 UK general election~\cite{curtis2017how}.
\emph{TheGuardian.com} readers tend to upvote comments about their preferred party more often than comments about the opposite Tory party. 
This bias exemplifies why upvote counts cannot readily be used to distinguish high-quality from low-quality comments.
Upvotes are cast with a subjective opinion in mind rather than with an objective and unbiased view of the comment text only.
Another interesting example are comments that mention the word \emph{people}.
These comments receive few upvotes but many replies, probably because they make generalized claims about groups of people, which are controversial and serve as conversation starters.
They are comparably unpopular on the platform but trigger many disapproving replies.
Two examples are: ``The people who voted for the war should be sent to prison as well.'' and ``People are disillusioned with mainstream politics, and are starting to look elsewhere.''.

\subsection{Taxonomy of Engaging Comments}
\begin{figure}
\centering
\includegraphics[width=1\columnwidth]{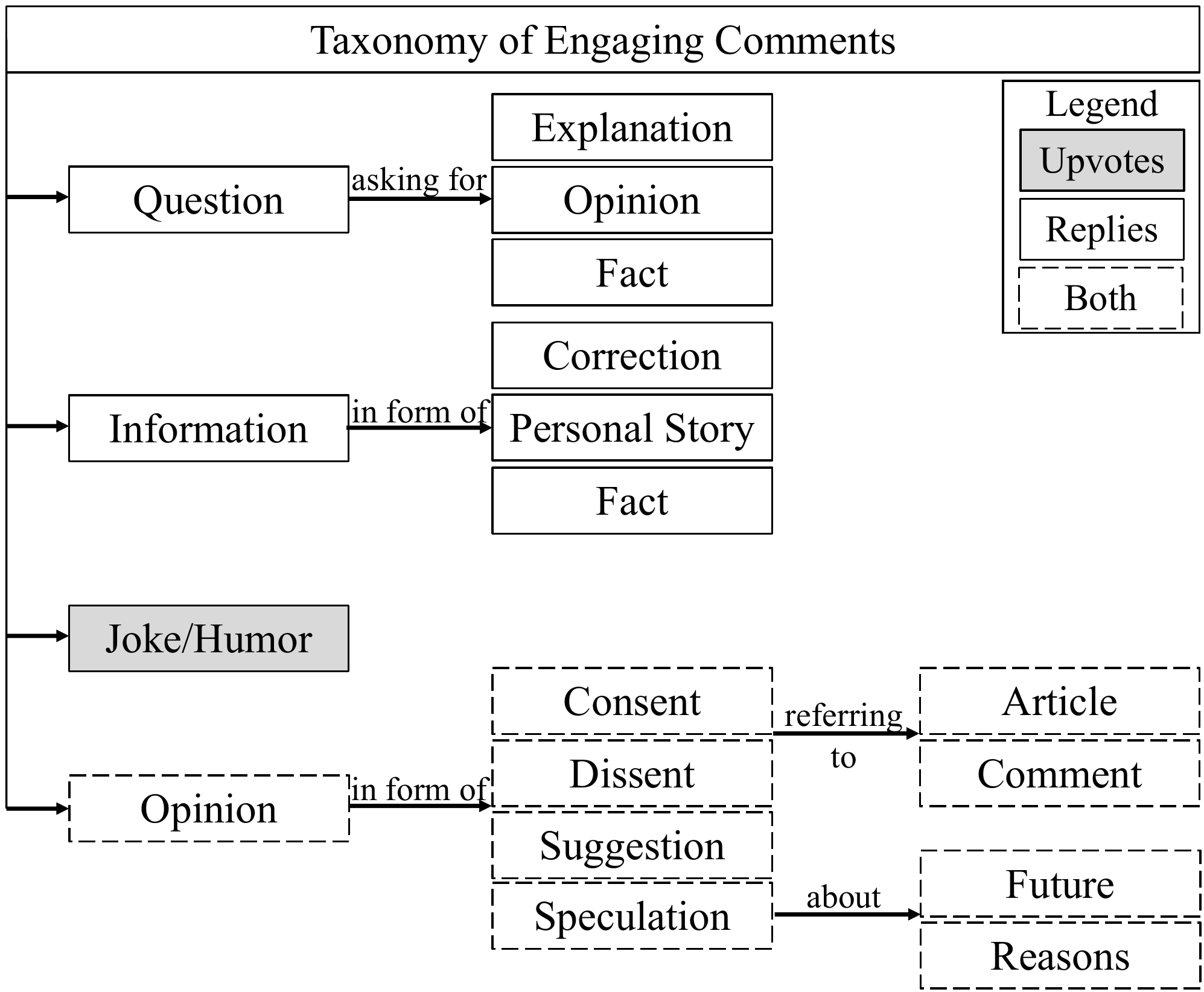}
\caption{Our hierarchical taxonomy of engaging comments classifies comments that attract upvotes (grey fill), replies (solid outline) or both (dashed outline).}
\label{fig:taxonomy}
\end{figure}

\begin{figure}[hbt!]
\centering
\includegraphics[width=1\linewidth]{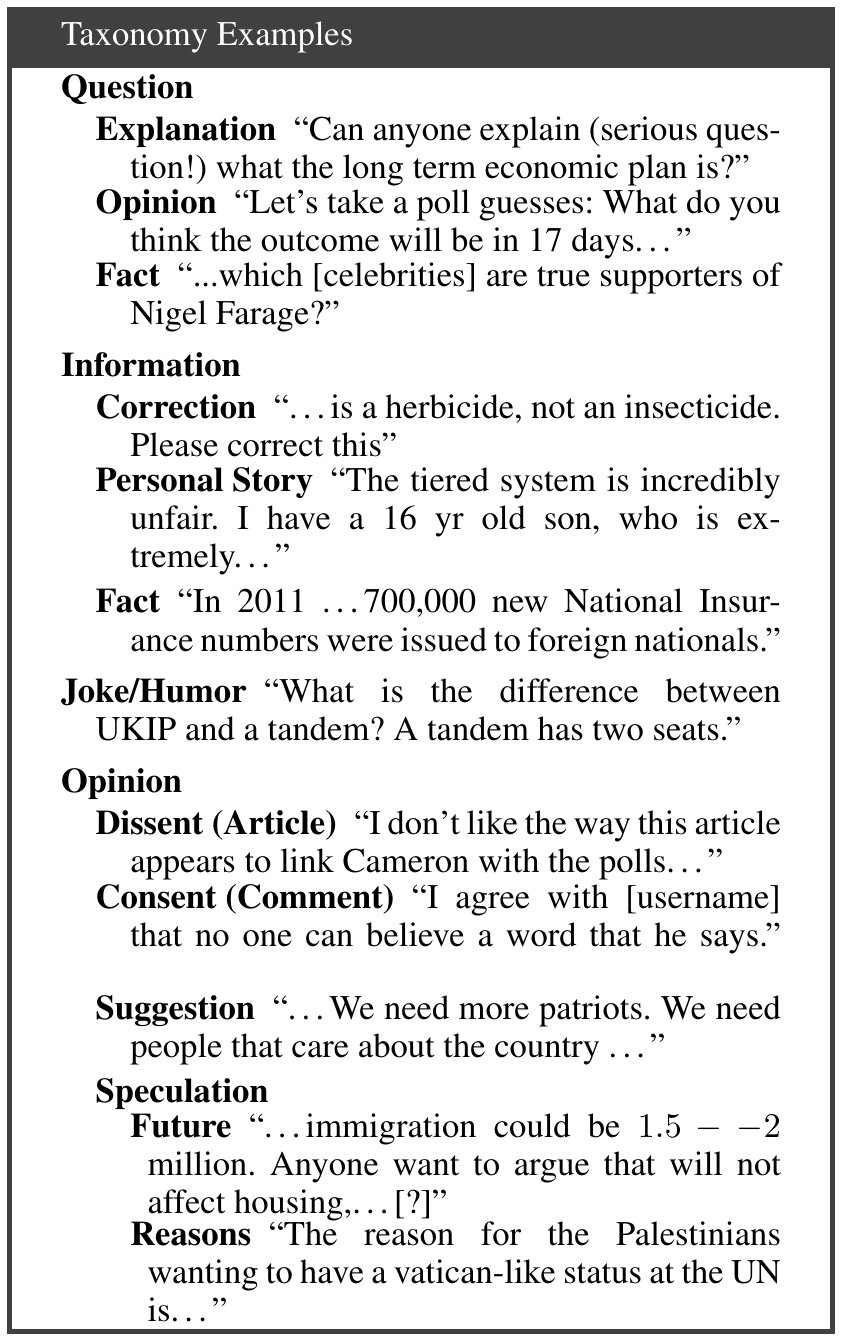}
\caption{A list of comments exemplifying each class in our hierarchical taxonomy of engaging comments (Figure~\ref{fig:taxonomy}).}
\label{fig:taxonomy-examples}
\end{figure}

Different taxonomies have been proposed for hateful comments~\cite{salminen2018anatomy,waseem-etal-2017-understanding} but not for engaging comments.
To foster a better understanding of engagement triggers, we propose a taxonomy for engaging comments, which is shown in Figure~\ref{fig:taxonomy}.
We follow an open coding approach, also used by \citeauthor{salminen2018anatomy}~\shortcite{salminen2018anatomy}, and code 1500 engaging comments. 
With this approach, we organize classes in a conceptual hierarchy.
Figure~\ref{fig:taxonomy-examples} exemplifies each class with a sample comment. 
For example, the class \emph{Question} groups the subclasses \emph{Explanation}, \emph{Opinion}, and \emph{Fact} together because all of them generate engagement by requesting answers in the form of comment replies.
Comments in all three subclasses typically contain an exclamation mark.
Note that the example comments for other classes, such as \emph{Joke/Humor} and \emph{Speculation} also contain questions.
However, these questions are more of a rhetorical nature, and the corresponding comments trigger engagement for other reasons.
The taxonomy also distinguishes between comments that trigger only upvotes, replies, or both.
For example, while comments with jokes rarely receive replies, they frequently receive upvotes.
It is the opposite if a comment asks for other users' opinions.
However, if a comment dissents from a news article, other users express their approval or disapproval with both upvotes and comments.
Our taxonomy is constructed in particular for comments on \emph{TheGuardian.com} and is by no means universal. 
Other platforms might exhibit other classes of engaging comments, for example, if they allow users also to downvote comments.
We revisit our taxonomy in the evaluation to understand which types of engaging comments are especially challenging to detect automatically.

\section{Distinguishing Top and Flop Comments}
\label{sec:approach}
We present a neural network model to distinguish top and flop comments based on their text.
Instead of labeling comments in a time consuming process, we draw upon the number of upvotes and replies that a comment received.
To this end, we consider only the comment text and remove the bias of the comment's rank and the news article topic.
We build two datasets of top and flop comments, which we then use to train our model with supervised learning.

\subsection{Removing the Position Bias}
We assume that a comment receiving many upvotes or replies is relevant to many users, whereas a comment with no or only a few reactions is comparably irrelevant.
However, this assumption only holds if malicious intentions to manipulate the user votes, e.g., voting multiple times with fake user accounts, can be ruled out.
In our dataset with 260 million upvotes from \emph{TheGuardian.com}, there is no incentive for users to manipulate the upvote count because it does not influence the order in which comments are displayed.
Occasional hoax upvotes can be neglected and considered noise.     
This leaves us with the vast majority of upvotes actually presenting an engagement signal on the level of individual comments.
Still, the number of upvotes and replies is biased by a comment's visibility to readers, which is influenced by the article's popularity and the comment's rank.

News platforms sort comments chronologically and show only the first few (e.g., ten) comments to readers directly below an article text.
All subsequent comments are hidden by pagination, which the user can access by browsing to the next page.
In practice, most users access only the very first page, which by default shows the oldest comments.
They never read any subsequent comments.
For this reason, we consider only the first ten comments directly below each article, ensuring that they were seen (and judged) by many readers. 
Articles with fewer than ten comments are discarded to allow for a fair comparison.

Some news articles draw more attention than others.
Thus, they attract a varying number of users who eventually consider voting on and replying to comments.
To normalize this variation, we transform the absolute number of upvotes and replies into relative numbers within each article's comment section.
To also remove the position bias illustrated in Figure~\ref{fig:upvotes_and_replies_by_rank}, we group all comments across all articles by their rank.
The result comprises ten groups of equal size.

We sort the comments of each rank by the descending relative number of upvotes.
Each sorted list now contains the comments in a normalized way.
All comments in the top 50 percent of the list perform better than an average comment at this rank, which means they received a comparably large portion of upvotes.
All comments in the bottom 50 percent received fewer upvotes than an average comment at this rank.
Thereby, the list contains top comments and flop comments with regard to upvotes, which can be used as positive and negative training samples for supervised learning.

There is only one variation for processing the replies. 
Articles that received less than 20 replies on their first ten comments are discarded.
In the same way as before, we then sort the comments of each rank by the descending relative number of replies.
Splitting each list in halves, results in sets of comments that receive more or fewer replies than an average comment at the respective rank.

By further filtering the dataset, e.g., to only the top 10 percent and bottom 10 percent, we consider only comments that perform much better or much worse than average.
This step can be seen as a way to filter for a higher agreement on a comment's rating among users.
Typically, upvotes and replies exhibit a low agreement: Users do not agree on which comments deserve upvotes or replies.
However, the agreement in the top 10 percent and bottom 10 percent is higher by definition of this subset of the data.
A much larger, respectively much lower, relative number of users reacted to the comments in these smaller sets.

Given the positive and negative training samples for supervised learning, we describe the architecture of our neural network model.
While we train two separate models, one for upvotes and one for replies as the measure to distinguish top and flop comments, the models have the same architecture.
We propose a recurrent neural network model based on Gated Recurrent Units (GRUs)~\cite{cho2014learning}.
The network starts with a pre-trained word embedding layer with fixed weights.
We pre-train 300-dimensional word embeddings on our full dataset of 61 million comments.
More precisely, we use the skip-gram training method of the fastText algorithm, which allows for mapping even out-of-vocabulary words to embedding vectors~\cite{bojanowski2016enriching}.
4.4 billion tokens are processed, which is about the same number of tokens as in the English Wikipedia.
The full text is lowercased and user mentions and URLs are replaced with special tokens.
We use the standard size of subwords of 3 to 6 characters and train for 5 epochs.
The same pre-trained word embeddings are used for both tasks and thereby the learned word representations are shared across them.

The second layer is a spatial dropout layer, which discards a fraction of the input words for regularization purposes.
It is followed by a layer of bidirectional Gated Recurrent Units (GRUs).
The bidirectionality allows each unit to consider both previous and subsequent units as context~\cite{schuster1997bidirectional}.
The output of the GRU layer passes a dropout layer and a dense layer.
A dense layer with a softmax activation and two outputs handles the final classification.
The network is trained with the Adam optimizer and binary cross-entropy as the loss function.

\section{Experiments}
\label{sec:experiments}
The first experiment evaluates the classification accuracy on a dataset of comments from \emph{TheGuardian.com}.
We compare four classifiers: (1) logistic regression on text length (baseline); (2) logistic regression on text and user features~\cite{park2016supporting}; (3) a convolutional neural network (CNN)~\cite{kim2014convolutional}; and (4) our recurrent neural network based on gated recurrent units (GRU).
Second, we use explanation methods for neural networks to investigate which words have the strongest influence on our model's predictions.
Finally, we evaluate classification accuracy on another dataset, which consists of product reviews from \emph{Amazon.com}.

\subsection{User Comments}
We consider the task of classifying comments into the classes \emph{top} and \emph{flop 10 percent} with regard to the normalized relative number of upvotes or replies received.
For example, a comment classified as \emph{top 10 percent} received a larger relative number of upvotes than 90 percent of the comments with the same rank.
We use classification accuracy as the evaluation metric because of the balanced class distribution.

To train the GRU-based model, early stopping on the decrease of validation loss determines the number of training epochs.
We set the number of neurons for the GRUs to 32, the dropout to 0.1, and the number of neurons of the dense layer to 16, and refrain from extensive hyperparameter optimization.
For comparison, we implement two state-of-the-art approaches: a CNN for sentence classification by \citeauthor{kim2014convolutional}~\shortcite{kim2014convolutional} and a feature-based classification approach by \citeauthor{park2016supporting}~\shortcite{park2016supporting}.
Kim's CNN uses a single layer of convolutions and max-over-time pooling.
Due to the relatively small number of parameters in this layer, the emphasis is put on the word embedding layer.
The feature-based classification approach by \citeauthor{park2016supporting}~\shortcite{park2016supporting} was specifically developed to support moderators in identifying high-quality online news comments.
It uses the following features:
comment length, comment readability, average comment length per user, average comment readability per user, and average number of received comment upvotes per user. 
These features serve as the input for a logistic regression classifier.
In addition to the two approaches from related work, we consider a naive baseline: a logistic regression classifier based solely on comment length.

\paragraph{Results}
Table~\ref{tab:experiment} shows the accuracy on the task of classifying top and flop comments with regard to upvotes and replies.
The column with the name \emph{10} refers to training on a dataset with the two classes \emph{top 10 percent} and \emph{flop 10 percent,} which contains 20k comments.
There are two more variants of the experiment also listed in Table~\ref{tab:experiment}.
The column with the name \emph{25} refers to training on a dataset with the two classes \emph{top 25 percent} and \emph{flop 25 percent,} which contains 53k comments.
The column with the name \emph{50} refers to training on a dataset with the two classes \emph{top 50 percent} and \emph{flop 50 percent,} which contains 106k comments.
While the training data differ, we use a shared test dataset split from the top/flop 10 percent because the labels in this dataset are the most reliable.
The other datasets contain more samples but are noisier.
The remaining data for each variant are split into 80 percent training and 20 percent validation.
We make sure that there is no overlap between the shared test set and any of the training and validation datasets.
Each experiment is repeated ten times. 
\begin{table}
  \caption{Classification accuracy on the task of distinguishing top and flop comments on \emph{TheGuardian.com.} with regard to the number of received upvotes and replies. 
  }
  \label{tab:experiment}
    \centering
  \fontsize{9.0pt}{10.0pt} \selectfont
  \begin{tabular}{lcccccc}
    \toprule
    & \multicolumn{3}{c}{Upvotes} & \multicolumn{3}{c}{Replies} \\
    \midrule
    Top/Flop \% & $10$ & $25$ & $50$ & $10$ & $25$ & $50$ \\
    \midrule
    Baseline & 0.61 & 0.61 & 0.61 & 0.63 & 0.63 & 0.63\\
    Park et al. 2016 & 0.65 & 0.66 & 0.67 & 0.61 & 0.59 & 0.60\\
    Kim 2014 & 0.67 & 0.63 & 0.62 & 0.69 & 0.65 & 0.67 \\
    Our Approach & \textbf{0.71} & \textbf{0.71} & \textbf{0.71} & \textbf{0.70}& \textbf{0.72} & \textbf{0.68} \\
\bottomrule
\end{tabular}
\end{table}

We perform a paired one-tailed t-test with a 95 percent confidence level to test the significance of our findings.
Our null hypothesis is that the true mean difference of the classification accuracy of the GRU and CNN approach is less than or equal to zero.
The null hypothesis is rejected for all our experiments, leaving us with strong evidence that the GRU approach outperforms the CNN approach with regard to classification accuracy.
The results in Table~\ref{tab:experiment} further show that the limitation to the top/flop 10 percent for training in general does not improve classification accuracy.
A consequence of this limitation are more reliable labels but also a smaller number of training samples. 
The GRU approach achieves the best performance on both tasks, upvote and reply prediction.
To our surprise, this approach, the logistic regression baseline on comment length only, and the feature-based approach of Park et al.\ are robust to the different variants of training data (top/flop 10, 25, 50 percent).
However, the CNN approach is less robust and performs better if trained on the top/flop 10 percent dataset.
If trained on the other dataset variants, the model overfits and does not generalize well to the test data.

\subsection{Explaining Predictions}
\citeauthor{arras2017explaining}~\shortcite{arras2017explaining} explain neural network predictions in the context of sentiment analysis.
We extend their approach to better understand what makes some comments more engaging than others.
To this end, we sort all words in the vocabulary according to their relevance for our model predicting \emph{high engagement} in the form of many upvotes or replies.
These word relevance scores are calculated with four methods: layer-wise relevance propagation (LRP)~\cite{bach2015pixel}, gradient-based sensitivity analysis (SA)~\cite{li2016understanding}, integrated gradients~\cite{sundararajan2017axiomatic}, and a random baseline.

The goal of the experiment is to measure how the deletion of different words changes the classification accuracy of our GRU model.
If we consider only true positives, the accuracy in this set is initially 1.
The accuracy decreases when we delete the words that are most relevant for the model's prediction and re-run the classification afterward.
If we consider only false negatives, the accuracy in this set is initially 0.
The accuracy increases when we delete the words that are least relevant for the correct class and re-run the classification.
The words that are deleted speak against the correct class.
Therefore, if their deletion changes the classification in favor of the correct class, accuracy increases.

\begin{figure}
\centering
\includegraphics[width=1\linewidth]{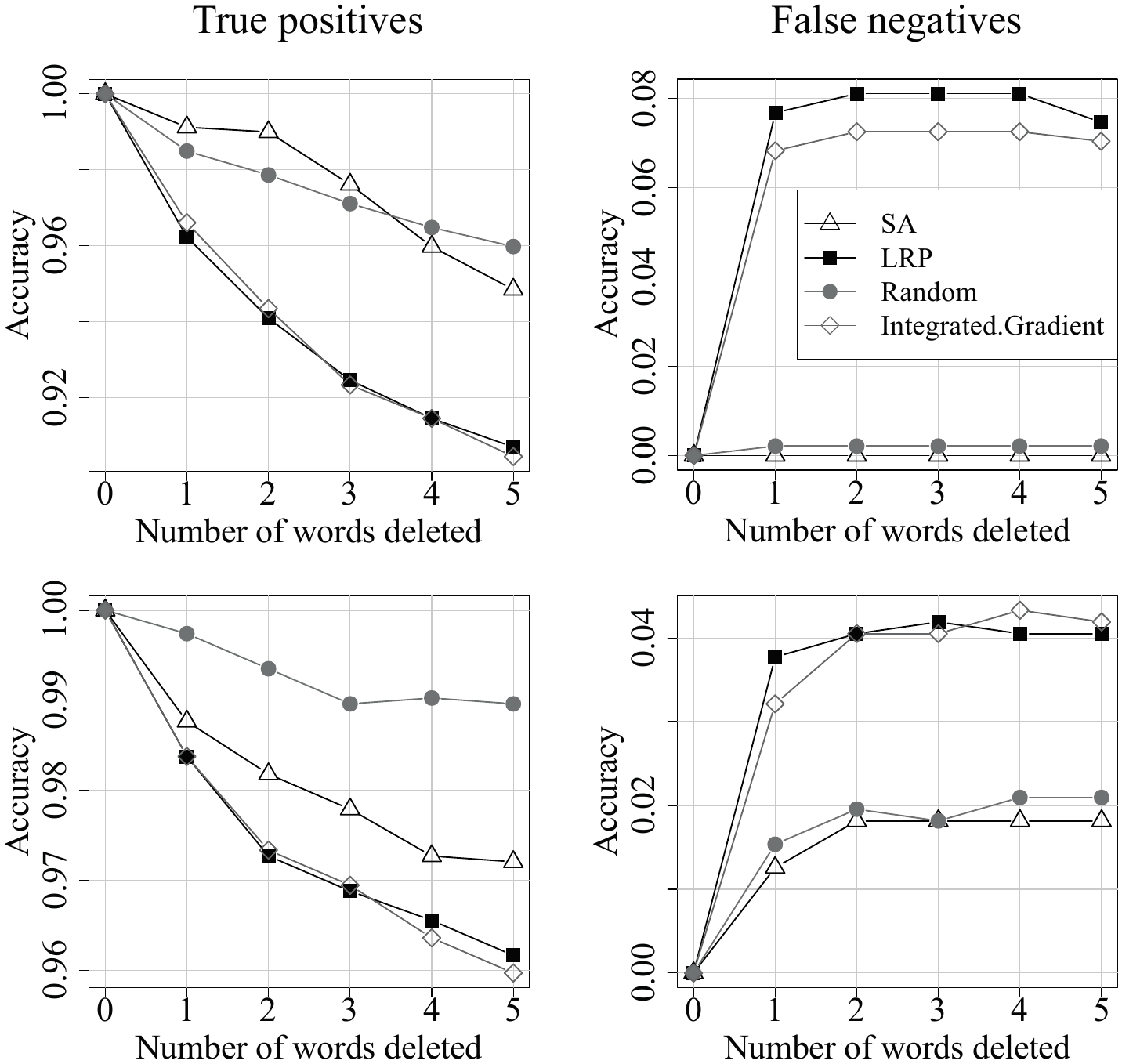}
\caption{Deleting the most relevant words from true positives (left-hand part) and the least relevant words from false negatives (right-hand part) has the strongest effect on reply (upper part) and upvote prediction (lower part) when using LRP relevance scores.}
\label{fig:word-deletion}
\end{figure}

Figure~\ref{fig:word-deletion} visualizes how deleting the most/least relevant words affects classification accuracy of our GRU model.
The larger the change in accuracy, the better are the calculated word relevance scores.
The two methods LRP and integrated gradients provide almost the same relevance scores and both outperform the method SA and the random baseline.

Based on layer-wise relevance propagation (LRP)~\cite{bach2015pixel}, we identify the most and least relevant words for our model's decisions.
Words that refer to strong emotions or controversial topics \emph{(arrogant, depressing, fantastic, bearable, Brexit)} are most relevant for predicting upvotes.
Least relevant are stop words \emph{(won't,} \emph{wasn't)} or emotions that are typically expressed in short comments \emph{(lol,} \emph{sigh)}.
Most relevant for predicting many replies are words referring to the Labour party \emph{(socialist,} \emph{lefty)}, which corresponds to the political orientation of most \emph{TheGuardian.com} readers.
The least relevant words are names of British public figures \emph{(Pickles,} \emph{Keir,} \emph{Tanner,} \emph{Morgan)}.

\begin{figure}
\centering
\includegraphics[width=1\linewidth]{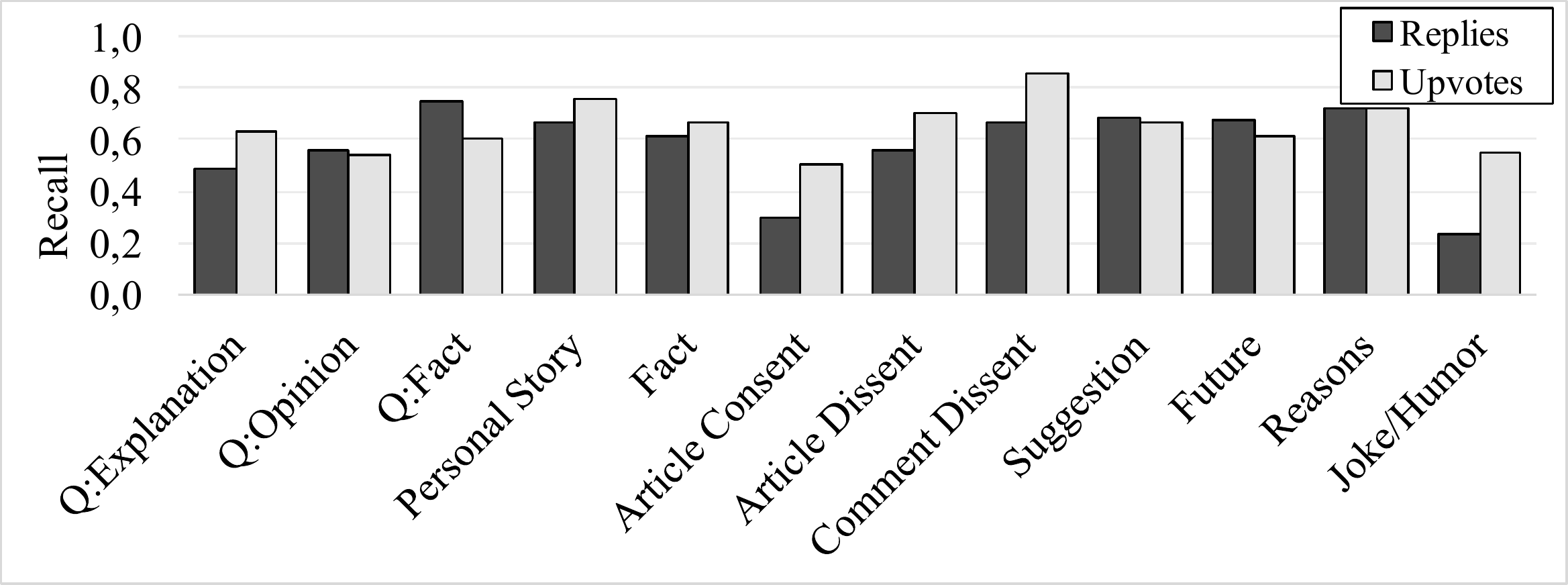}
\caption{Recall for identifying engaging comments differs per class, e.g., jokes are less reliably identified than dissent.}
\label{fig:labeled-test-data}
\end{figure}

According to our taxonomy for engaging comments, we labeled all positive samples in the test set of the \emph{top/flop 10 percent} dataset.
For each class, Figure~\ref{fig:labeled-test-data} shows our model's recall at distinguishing top and flop comments.
The classes \emph{Correction} and \emph{Comment Consent} are omitted because there was only a handful of such samples in the test set.
The recall for \emph{Joke/Humor} is lowest, whereas the recall for \emph{Comment Dissent} or speculation about \emph{Future} and \emph{Reasons} is highest.
This discrepancy means that our model's predictions could be improved by a better detection of \emph{Joke/Humor}.
Further, questions asking for facts (\emph{Q:Fact}) are identified with higher recall than comments providing facts (\emph{Fact}).
Besides these differences, the recall for all classes is similar.

\subsection{Product Reviews}
User comments on news platforms and product reviews on online retail platforms have several properties in common:
(1) popular news articles, as well as popular products, generate an overwhelming number of posts, (2) posts on both platforms are typically short, and (3) both allow users to vote on posts.
On product review platforms, upvotes resemble votes on the helpfulness of a review.
However, news discussions differ from disconnected posts on Amazon, YouTube, and Twitter, where no discussions take place and the communication is unidirectional.
Reviews focus on a particular product and do not refer to each other.
\citeauthor{danescu2009how}~\shortcite{danescu2009how} analyze a dataset of Amazon product reviews and their helpfulness votes.
They find that users consider a review more helpful if the associated product rating is closer to the average rating for this product (conformity bias).

We consider product reviews posted on \emph{Amazon.com} to study the applicability of our approach to other domains.
The dataset contains 82 million Amazon product reviews, spanning from May 1996 to July 2014, and is available online~\cite{he2016ups}.
170 million upvotes (``Was this review helpful?'') were cast in total.
We filter the dataset so that we consider the ten earliest reviews per product.
Products with less than ten reviews are discarded.
Similar to the dataset of user comments at \emph{TheGuardian.com}, we learn word embeddings on this large dataset.
The reviews comprise 7.6 billion tokens, which is more than twice the number of tokens in the English Wikipedia.

For a classification experiment, we use a subset of 9 million book reviews to reduce topical variety.
We apply the same normalization steps to upvote counts as described earlier and consider three different variants of the dataset.
They correspond to the top and flop 10, 25, and 50 percent of the product reviews and contain 220k, 550k, and 1.1 million product reviews, respectively.
The test set is shared for all variations of the training data and comprises 10 percent of the \emph{top/flop 10 percent} dataset (22k reviews).
The remaining data for each variant are randomly split into 80 percent training and 20 percent validation set.

We compare the classification accuracy of the logistic regression baseline on review length, the CNN by \citeauthor{kim2014convolutional}~\shortcite{kim2014convolutional}, and our approach on the task of distinguishing helpful (top) and non-helpful (flop) product reviews.
The feature-based classifier by \citeauthor{park2016supporting}~\shortcite{park2016supporting} cannot be applied to the product reviews because it requires user information, which our dataset does not contain.

\paragraph{Results}
Table~\ref{tab:experiment2} lists the results of the experiment. 
The GRU model outperforms the CNN.
In contrast to our comment dataset, the limitation to the top/flop 10 percent on the product reviews dataset improves classification accuracy.
Here, the training dataset is ten times larger, which diminishes the disadvantage of limiting the data to the top and flop 10 percent.
The more reliable labels in the top/flop 10 and 20 percent training datasets make the difference.
\begin{table}
  \caption{Classification accuracy on the task of distinguishing top and flop product reviews on \emph{Amazon.com} with regard to the number of received helpfulness upvotes.}  
  \label{tab:experiment2}
    \centering
  \begin{tabular}{lccc}
    \toprule
    Top/Flop Percent & $10$ & $25$ & $50$ \\
    \midrule
    Baseline & 0.67 & 0.67 & 0.34 \\
    Kim 2014    & 0.67 & 0.72 & 0.64 \\
    Our Approach         & \textbf{0.76} & \textbf{0.75} & \textbf{0.66} \\
  \bottomrule
\end{tabular}
\end{table}

Training on the top/flop 50 percent dataset results in the worst performance.
For the baseline that considers only the comment length, it is even worse than random guessing, which achieves 50 percent accuracy.
The different value distributions of review lengths in training and test data explains this result.
The baseline is unable to learn an appropriate threshold for the comment length.
The most and least engaging product reviews in the top/flop 50 percent dataset have similar average length (1076 vs.\ 1055 characters), whereas there is a clear separation for review lengths in the top/flop 10 percent dataset (677 vs.\ 1387 characters).

\section{Impact on Online Discussions}
\label{sec:impact}
Many online news platforms closed their comment section under the unbearable workload of content moderation and hateful and abusive comments.
But, hateful comments and abusive language are not the only problem.
Without in-depth discussions, where users exchange reasonable arguments for their opinions, comment sections create (almost) no added value for the news platforms.
Users shout out their own opinions but rarely ever listen to each other and start fruitful conversations.
To stand out among the plethora of online spaces where users can post their opinions, modern news platforms need to add value by providing a space for engaging and polite exchange.
Any change to comment sections that fosters user interaction or increases commitment by design~\cite{aragon2017thread,farzan2011increasing,budak2017threading} can make a big difference.

Today, comments in online discussions are mostly ranked chronologically --- with a few exceptions, such as \emph{Slashdot.org} and \emph{Digg.com}.
While one might argue that this approach is transparent and fair, it does not foster engaging discussions.
Instead, it only gives an incentive to post comments as fast as possible after an article is published.
In that case, the comment will get ranked high, it will gain visibility in the community, and possibly get some reactions to the comment, no matter how good or bad it is.
This competition goes so far that some users refrain from reading the article to be the first to post a comment.
Our approach introduces an alternative method to rank comments by the expected number of upvotes and replies.
If applied, on the one hand, the visibility of top-performing, engaging comments increases.
They are shown to more users.
On the other hand, the least engaging, flop comments lose visibility and are practically hidden at the end of the comment section, which usually no user accesses.

A limitation of our study is that we only consider a comment's text content and no user-based features.
The reputation of the comment author presumably affects its impact in terms of visibility and thus received upvotes and replies.
Further, the most comment texts that we explored are well-formed and grammatically correct, which simplifies the analysis.
Emoticons and slang are rarely used on \emph{TheGuardian.com}.
However, they might be more frequent on other platforms and pose a potential challenge.
Design changes on the online platforms are an additional challenge for analyzing a long timespan.
For example, with the most recent features, users can sort comments by time or by the number of upvotes. 
The default setting of the sorting, e.g. newest/oldest first, is an important factor for the visibility of individual comments.
Editor picks change the visibility of selected comments in a similar way. 

\section{Conclusions and Future Work}
\label{sec:conclusions}
We studied comment upvotes and replies as a measure of user engagement in online news discussions.
To this end, we designed a taxonomy of engaging comments on the platform \emph{TheGuardian.com} and analyzed textual differences of the most and least engaging comments (top and flop comments).
Further, we trained a neural network model to distinguish these top and flop comments given the comments' texts.
To construct the training dataset, we identified and removed the position bias that favors early comments and normalized upvote counts for each article individually to also remove a potential topical bias.

Based on predicted user reactions in the form of upvotes and replies, platforms could automatically highlight or rank comments to show top ones to users and thus encourage more interaction.
Experimental results demonstrate that neural network models outperform feature-based classification approaches and achieve an accuracy of about 70 percent on a balanced test dataset.
This result is not limited to our dataset of comments and also generalizes to product reviews.

A promising path for future work is to investigate different types of votes, e.g., not only upvotes but also downvotes or more fine-grained votes.
It would be interesting to analyze the interplay of these types.
A comment that receives many upvotes and downvotes at the same time might be considered controversial.
Another idea is to consider user names and reputation as a predictive feature in the classification process. 
For example, a comment by a journalist might generate more engagement than a comment by a regular user.

\bibliographystyle{aaai}
\bibliography{bibliography-ready}
\end{document}